\documentclass[aip,apl,preprint,groupedaddress,showpacs]{revtex4-1}
\usepackage[dvips]{graphicx}
\begin{document}


\title{Origin of intrinsic dark count in superconducting nanowire single-photon detectors}

\author{T. Yamashita}
\email[]{taro@nict.go.jp}
\author{S. Miki, K. Makise, W. Qiu, H. Terai, M. Fujiwara, M. Sasaki} 
\author{Z. Wang}
\affiliation
{Advanced ICT Research Institute, National Institute of 
Information and Communications Technology (NICT), 588-2 Iwaoka, Kobe 651-2492, Japan}

\date{\today}

\begin{abstract}
The origin of the decoherence in superconducting nanowire single-photon detectors, 
the so-called dark count, was investigated. 
We measured the direct-current characteristics and bias-current dependencies of the dark count rate 
in a wide range of temperatures from 0.5 K to 4 K, and analyzed the results 
by theoretical models of thermal fluctuations of vortices. 
Our results indicate that the current-assisted unbinding of vortex-antivortex pairs is 
the dominant origin of the dark count. 
\end{abstract}


\maketitle

Single-photon detectors are recently attracting a great deal of attention in a wide range of fields, 
such as single-photon source characterization, optical communication, and quantum information. 
\cite{Hadfield1,Robinson,Sasaki} 
Among the various types of the single-photon detectors, the superconducting nanowire single-photon detector 
(SNSPD), which consists of a meander line of ultrathin superconducting nanowire, 
is promising because of its many advantages. \cite{Goltsman,Hadfield2} 
In fact, the SNSPDs are incorporated in a quantum key distribution network over a long distance 
and play an essential role for the realization of quantum cryptography. \cite{Sasaki}

One important feature of the SNSPD is a low dark count rate (DCR). \cite{Hadfield2} 
The DCR is suppressed as the bias current is reduced, but 
the detection efficiency (DE) is also simultaneously reduced. 
In order to obtain a high DE, one should apply the large bias current near the critical current. 
It is known that the dark counts in the high bias region remains finite even by shielding the device 
from external noise and are dominated by the {\it spontaneous} resistive-state formation 
which is a decoherence phenomenon intrinsic to thin-film superconducting nanowire. 
Therefore it is vital to clarify the origin of the intrinsic dark counts to establish means 
of reducing the DCR and achieve the full potential of the devices. 

To date, the mechanism of the dark count has been studied, but not clearly identified. 
\cite{Engel,Kitaygorsky,Bartolf,Bulaevskii} 
Refs. 6 and 7 discussed the possibility of vortex-antivortex pair (VAP) unbinding 
due to Berezinskii-Kosterlitz-Thouless (BKT) transition, \cite{Berezinskii,Kosterlitz} 
and suspected that VAP unbinding is the likely source but needed further investigation. 
Another possible mechanism is the vortex hopping from an edge of the nanowire. \cite{Bartolf,Bulaevskii} 
Recently, Bartolf {\it et al.} investigated the DCR precisely 
by using several fluctuation models of vortices. \cite{Bartolf} 
However, the two possible mechanisms, 
VAP unbinding and vortex hopping, have not been discriminated because they both give 
the same order of the DCR at the temperature of around 5.5 K that they presented. \cite{Bartolf} 
In order to conclusively identify the dominant origin of the dark count in the SNSPD, 
measurements of the wide-range temperature dependencies of the DCR 
and the direct-current (DC) characteristics are required. 

In this letter, we measured the current-voltage and resistance-temperature 
characteristics of the SNSPD and the bias-current dependencies of the DCR at several temperatures 
from 0.5 K to 4.0 K, and analyzed the results by using theoretical models to identify the origin 
of the intrinsic dark count in the SNSPD. 


The SNSPD device presented in this letter consists of NbN film with a thickness $d$ of 4 nm, 
which was grown on a single-crystal MgO substrate by reactive DC magnetron sputtering. 
The NbN thin film formed a meander line with an area of 20 $\mu$m $\times$ 20 $\mu$m. 
The nanowire width $w$ = 100 nm and the pitch is 60 nm leading to the filling factor of 62.5\%. 
The critical current density $J_{c}$ was $3.1 \times 10^{10}$ A/m$^{2}$ at 3.0 K. 
The penetration depth $\lambda(0)$ = 495 nm was evaluated from the superconducting transition temperature 
and the resistivity in the normal state. \cite{Wang} 
The Ginzburg-Landau (GL) coherence length is typically at zero temperature 
$\xi(0)$ = 5 nm for NbN thin film. 
The fabrication process of SNSPDs has been detailed elsewhere. \cite{Miki_IEEE} 


SNSPD devices can be treated as a two-dimensional (2D) superconductor since 
the thickness of the nanowire is comparable to the coherence length 
($d\approx\xi$) whereas the width is much larger ($w \gg \xi$). 
In the 2D superconductors, it is known that a phase transition called the BKT transition can 
occur at the critical temperature $T_{\rm BKT}$. \cite{Berezinskii,Kosterlitz} 
Below $T_{\rm BKT}$, the formation of vortices and antivortices is 
favored energetically and the VAP is spontaneously generated. 
In the SNSPD, the bias current is applied to the device and therefore the vortex and antivortex 
of a VAP feel Lorentz force in the opposite direction. 
As a result, the VAP is broken into single vortices easily compared to the case of no bias current. 
When the unbound vortices move in the nanowire perpendicular to the current, the finite resistance 
appears even below the superconducting critical temperature $T_{c}$; 
thus the current-assisted VAP unbinding can be a possible origin of the dark counts. 
\cite{Engel,Kitaygorsky,Bartolf} 
Regarding free vortices due to the finite-size effect below $T_{\rm BKT}$, 
their contribution to the dark count is expected to be small 
in the high bias-current region. \cite{Bartolf} 
Another possibility is dissipation due to the vortex hopping in the nanowire by 
the magnetic self field of the applied bias current. \cite{Maksimova} 
When the vortex generated due to the self field overcomes the edge barrier and crosses 
the nanowire by thermal excitation, decoherence appears. 
Regarding the single vortices induced by an external magnetic field, we can exclude them in our case 
because the device is shielded magnetically in the refrigerator. 
Thus, in the following discussion we consider two mechanisms: 
(i) current-assisted VAP unbinding and (ii) thermal-excited vortex hopping 
by the self-magnetic field of the bias current. 


Before entering upon a discussion of the dark count using the above models, 
we need to check whether the BKT transition occurs in the SNSPD or not. 
In order to confirm the BKT transition, 
we measured the temperature dependencies of the current-voltage characteristics 
and the resistance. \cite{Kadin,Mooij} 
We used a cryogen-free refrigerator for detailed measurements of 
the DC characteristics at temperatures from 4 K to 20 K. 
The temperature was well controlled by a proportional-integral-derivative controller. 
A $\mu$-metal magnetic shield surrounds the device-mounted stage in the refrigerator. 
Measurements were performed using a standard four-terminal setup. 

Figure 1(a) shows the current-voltage characteristics at temperatures 
between 8.40 K and 9.00 K. 
The BKT critical temperature $T_{\rm BKT}$ is determined as the temperature 
corresponding to $\alpha=3$ in the power-law relation $V \propto I^{\alpha}$. \cite{Kadin,Mooij} 
Figure 1(b) indicates the temperature dependence of $\alpha$ derived from 
the current-voltage characteristics. 
As shown in the figure, the value of $\alpha$ decreases with increasing the temperature, 
and reaches the value of $\alpha \approx 3$ at 8.50 K. 
From 8.50 K to 8.55 K, the value of $\alpha$ shows the jump 
which indicates that the BKT transition occurs in the nanowire, 
and thus we found $T_{\rm BKT}$ = 8.50 K. 
At the superconducting transition, the value of $\alpha$ becomes 1; $T_{c}$ = 9.0 K. 
The broaden transition of $\alpha$ from 3 to 1 has been reported in the finite-size superconductors. 
\cite{Gorlova,Herbert}
Beasley, Mooij, and Orlando (BMO) derived a relation expressed as 
${T_{\rm BKT}} = {T_{c}} \left( 1 + 0.173\epsilon_{c}{e^{2}R_{\rm SN}}/{\hbar} \right)^{-1}$ 
where $R_{\rm SN}$ is sheet resistance in the normal state, $e$ is elementary electric charge, 
and $\hbar$ is the reduced Plank constant. \cite{Beasley,Mooij} 
Here, $\epsilon_{c}$ indicates the vortex dielectric constant at $T_{\rm BKT}$ 
which depends on the density of VAPs located between the given test vortices. 
By substituting the values of $T_{c}$, $T_{\rm BKT}$, and $R_{\rm SN}$ of 648 $\Omega$ 
into BMO expression, we estimate $\epsilon_{c}$ = 2.15. 

Figure 1(c) shows the temperature dependence of the sheet resistance $R_{\rm S}$. 
It is indicated that the film is nominally homogeneous because there is no indication of 
the quasi-reentrant behavior of granular ultrathin films. \cite{Jaeger} 
The sheet resistance for $T_{\rm BKT} < T < T_{c}$ is expressed as 
$R_{\rm S}(T) = a R_{\rm SN} \exp\left( -2\sqrt{b({T_{c}-T})/({T-T_{\rm BKT}})} \right)$ 
where $a$ and $b$ are fitting parameters. \cite{Mooij} 
We fitted this equation to our data by using $T_{c}$ of 9.0 K derived from the current-voltage 
characteristics, and found $T_{\rm BKT}$ = 8.53 K, which shows good agreement with 
the value of 8.50 K obtained by the current-voltage measurement. 
From the above analyses on the current-voltage and resistance-temperature characteristics, 
it was confirmed that the BKT transition actually occurs in the SNSPD. 

Let us move on to the discussion on the origin of the dark count. 
We measured the bias-current dependencies of the DCR at temperatures of 0.5 K to 4 K. 
The DCR measurement was performed in a dilution refrigerator with a base temperature of 11 mK. 
The dilution unit is inside a refrigerator dewar mounted on a vibration-free stage. 
Double $\mu$-metal cylinders surround the dewar. 
The device was current-biased via the DC arm of the bias tee, and the output signal of the dark count 
was observed by the pulse counter through the AC arm of the bias tee and two low-noise amplifiers. 
Figure 2 shows the measured DCR as functions of the bias current $I_{b}$ 
normalized by the critical current $I_{c}$ at temperatures of 0.5 K to 4.0 K. 
The DCR increased with increasing the bias current at all temperatures, 
and decreases for lower temperatures.

The measurements of the DCR were performed below $T_{\rm BKT}$ and thus the vortices form 
the VAP in the nanowire by the BKT transition. 
The probability of the current-assisted VAP unbinding is expressed as 
$P_{\rm VAP} = \Gamma_{\rm VAP} \exp(-U_{\rm VAP}/k_{\rm B}T)$ with the attempt rate $\Gamma_{\rm VAP}$ 
and the Boltzmann constant $k_{\rm B}$. 
$U_{\rm VAP}$ is the potential at the saddle point expressed as \cite{Mooij}
\begin{eqnarray}
U_{\rm VAP} = 2\mu_{c} + \frac{A(T)}{\epsilon} \left[ \ln\left(\frac{2.6I_{c}}{I_{b}}\right) 
+ \frac{I_{b}}{2.6I_{c}} - 1 \right], 
\end{eqnarray}
where $\mu_{c}$ is the vortex core potential and $\epsilon$ is the dielectric constant depending on 
the density of the VAP. Here $A(T) = \Phi_{0}^{2}/\left(\pi\mu_{0}\Lambda(T)\right)$, 
where $\Phi_{0}$ is the flux quantum, $\mu_{0}$ the permeability, 
and $\Lambda(T) = 2\lambda(T)^{2}/d$ the effective penetration depth. \cite{lambda} 
In Fig. 2, the solid lines show the best-fitted curves with the experimental data 
at each temperature by using the least-square approach. 
We used the relation $\mu_{c} \approx \gamma A(T) - k_{\rm B}T\ln{N_{0}}$ with the temperature-independent 
constant $\gamma$ and a measure of the number of independent configurations $N_{0}$. \cite{Mooij} 
As shown in the figure, theoretical curves fit the data very well at all temperatures. 
In the fitting procedure, we obtained the values of $\epsilon = 1.4 - 6.5$ as a fitting parameter. 
The obtained values are reasonable in the framework of the BKT theory because 
they are comparable to $\epsilon_{c}$ of 2.15 derived from the BMO expression. 
Therefore, it is found that the current-assisted VAP unbinding model explains 
the experimental results well in the temperature range we measured. 

Another possible mechanism of the dark count is the vortex hopping overcoming the edge barrier. 
\cite{Bartolf,Maksimova} 
In Ref. 8, this mechanism contributed to the DCR by the same order as 
the VAP unbinding model at $T \approx 0.4 T_{c}$, and thus the dominant mechanism could not be concluded. 
The edge barrier potential as a function of the coordinate $x$ in the nanowire ($0<x<w$) 
is expressed as \cite{Bartolf,Maksimova} 
\begin{eqnarray}
U_{\rm VH}(x) = 
E_{B}(T) \Bigg\{ \ln\left[\frac{2w}{\pi\xi(T)}\sin\left(\frac{\pi x}{w}\right)\right] 
- \frac{I_{b}}{I_{B}(T)}\frac{\pi}{w}\left( x - \frac{\xi(T)}{2}\right) \Bigg\}, 
\label{eq:UVH}
\end{eqnarray}
where $E_{B}=\Phi_{0}^{2}/\left( 2\pi\mu_{0}\Lambda(T) \right)$, 
$I_{B}=\Phi_{0}/\left( 2\mu_{0}\Lambda(T) \right)$, and 
$\xi=\xi(0)/\sqrt{1-T/T_{c}}$ is the temperature dependent GL coherence length. \cite{Tinkham} 
The probability of the vortex hopping is expressed as 
$P_{\rm VH} = \Gamma_{\rm VH} \tilde{I}_{b} \exp(-U_{\rm VH,max}/k_{B}T)$ 
with the maximum point of the edge-barrier potential in the $x$-space $U_{\rm VH,max}$ which is 
derived from the equation $dU_{\rm VH}(x)/dx=0$, 
attempt rate $\Gamma_{\rm VH}$, and $\tilde{I}_{b}=I_{b}/I_{c}$. \cite{Bartolf} 
In Fig. 2, 
the dashed lines indicate the best-fitted curves by using the thermal-excited vortex hopping model. 
As seen in the figure, they did not fit to the experimental data at all temperatures. 
Note that only the curve at 2.5 K shows a similar dependence with the experimental data 
as reported in refs. 8 and 9, but the curves at the other temperatures 
deviate from the experimental data entirely. 
Therefore the thermally excited vortex hopping model could be excluded, and 
we conclude that the current-assisted unbinding of VAPs is the dominant origin of the dark count. 
In Ref. 9 which reached a different conclusion from the present letter, the authors mentioned that 
their vortex-hopping scenario explained the experimental data for samples on the sapphire substrate 
\cite{Bartolf} with the smaller width, but not for the large-width sample. 
In addition to measurements of the temperature dependence as performed in the present work, 
further investigations into the effects of the substrate and the sample width on the dark count 
will be valuable. 


In conclusion, we investigated the DC characteristics and bias-current dependencies 
of the dark count rate of a superconducting nanowire single-photon detector 
at temperatures of 0.5 K to 4.0 K. 
We identified that the current-assisted unbinding of vortex-antivortex pairs is 
the dominant mechanism of the dark count in the temperature range in which 
the SNSPDs are usually operated. 
Innovative device design for preventing the unbinding of the vortex-antivortex pairs and/or 
suppressing the vortex crossing will lead to the realization of dark-count-free SNSPDs. 

%






\newpage
{\bf FIGURE CAPTIONS}
\\
\\
Fig. 1: (a) Current-voltage characteristics of SNSPD at several temperatures. 
From right to left, temperatures are 8.40, 8.45, 8.50, 8.55, 8.60, 8.65, 8.70, 8.80, 8.90, and 9.00 K. 
(b) temperature dependence of $\alpha$ from 8.3 K to 9.0 K. 
(c) Temperature dependence of sheet resistance $R_{\rm S}$. 
\\
\\
Fig. 2: Bias-current dependencies of measured DCR at temperatures of 0.5 K to 4.0 K (symbols). The solid and dashed lines indicate best-fitted curves described by models of 
current-assisted VAP unbinding and thermal-excited vortex hopping, respectively, 
at 4.0 K$-$0.5 K from left to right. 

\newpage
\begin{figure}[t]
\begin{center}
\includegraphics{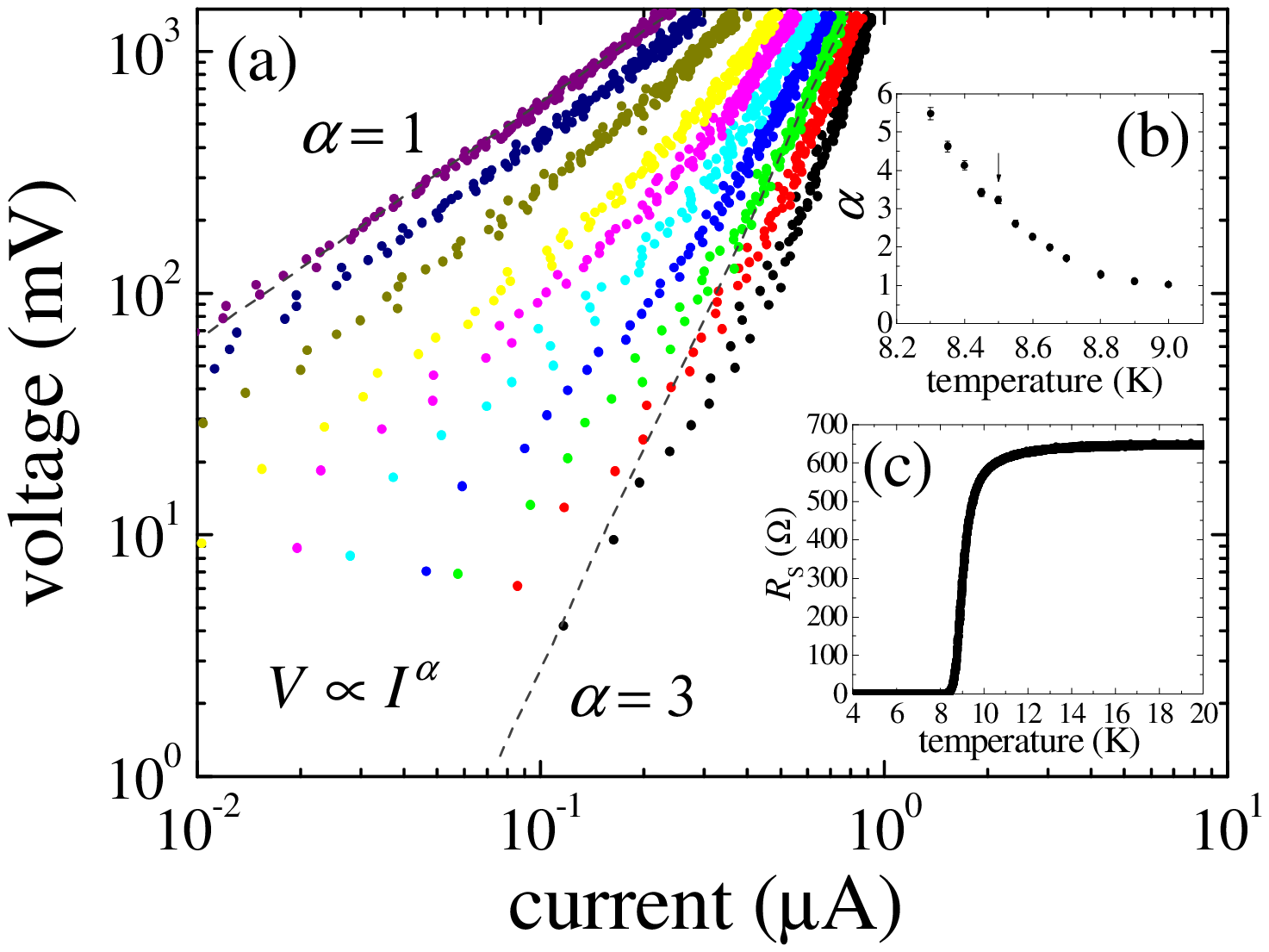}
\end{center}
\caption{\label{Fig:IV}}
\label{Fig:IV}
\end{figure}
\newpage
\begin{figure}[t]
\begin{center}
\includegraphics{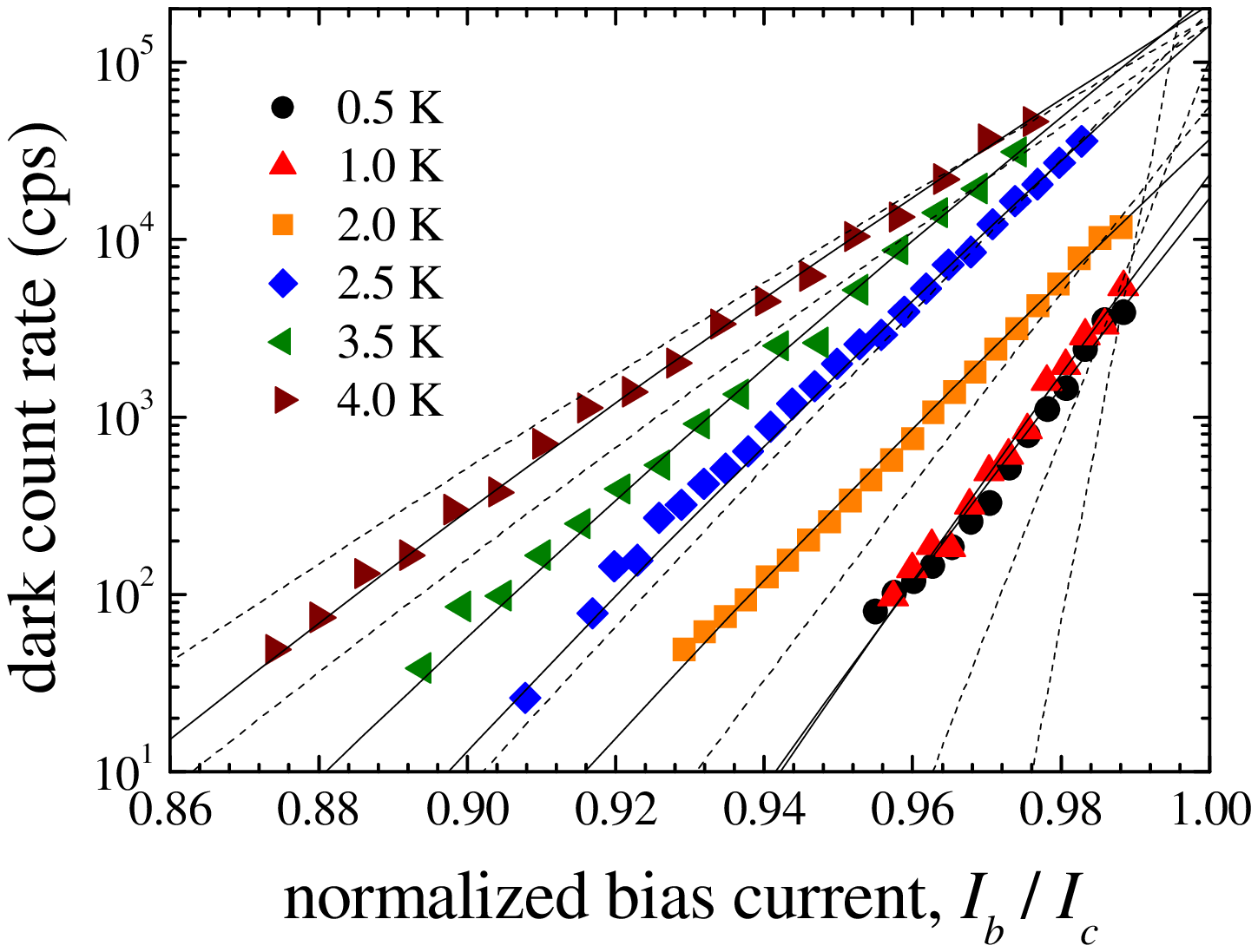}
\end{center}
\caption{\label{Fig:DCR}}
\label{Fig:DCR}
\end{figure}


\begin{thebibliography}{99}
\bibitem{Hadfield1} R.H. Hadfield, M.J. Stevens, S.S. Gruber, A.J. Miller, R.E. Schwall, 
R.P. Mirin, and S.W. Nam, Opt. Express {\bf 13} 10846 (2005).
\bibitem{Robinson} B.S. Robinson, A.J. Kerman, E.A. Dauler, R.J. Barron, D.O. Caplan, 
M.L. Stevens, J.J. Carney, S.A. Hamilton, J.K.W. Yang, and K.K. Berggren, Opt. Lett. {\bf 31} 444 (2006).
\bibitem{Sasaki} M. Sasaki, M. Fujiwara, H. ishizuka, W. Klaus, K. Wakui, M. Takeoka, S. Miki, T. Yamashita, 
Z. Wang, A. Tanaka, K. Yoshino, Y. Nambu, S. Takahashi. A. Tajima, A. Tomita, T. Domeki, T. Hasegawa, 
Y. Sasaki, H. Kobayashi, T. Asai, K. Shimizu, T. Tokura, T. Tsurumaru, M. Matsui, T. Honjo, K. Tamaki, 
H. Takesue, Y. Tokura, J.F. Dynes, A.R. Dixon, A.W. Sharpe, Z.L. Yuan, A.J. Shields, S. Uchikoga, M. Legire, 
S. Robyr. P. Trinkler, L. Monat, J.-B. Page, G. Ribordy, A. Poppe, A. Allacher, O. Maurhart, T. L{\"a}nger, 
M. Peev, and A. Zeilinger, Optics Express {\bf 19}, 10387 (2011). 
\bibitem{Goltsman} G.N. Gol'tsman, O. Okunev, G. Chulkova, A. Lipatov, A. Semenov, K. Smirnov, 
B. Voronov, A. Dzardanov, C. Williams, and Roman Sobolewski, Appl. Phys. Lett. {\bf 79}, 705 (2001). 
\bibitem{Hadfield2} R.H. Hadfield, Nature Photonics {\bf 3}, 696 (2009).
\bibitem{Engel} A. Engel, A.D. Semenov, H.-W. H{\"u}bers, K. Il'in, and M. Siegel, 
Physica C {\bf 444}, 12 (2006). 
\bibitem{Kitaygorsky} J. Kitaygorsky, I. Komissarov, A. Jukna, D. Pan, O. Minaeva, N. Kaurova, 
A. Divochiy, A. Korneev, M. Tarkhov, B. Voronov, I. Milostnaya, G. Gol'tsman, and R.R. Sobolewski, 
IEEE Trans. Appl. Supercond. {\bf 17}, 275 (2007). 
\bibitem{Bartolf} H. Bartolf, A. Engel, A. Schilling, K. Il'in, M. Siegel, H.-W. H{\"u}bers, and A. Semenov, 
Phys. Rev. B {\bf 81}, 024502 (2010). 
\bibitem{Bulaevskii} L.N. Bulaevskii, M.J. Graf, C.D. Batista, and V.G. Kogan, 
Phys. Rev. B {\bf 83}, 144526 (2011). 
\bibitem{Berezinskii} Z.L. Berezinskii, Sov. Phys. JETP {\bf 32}, 493 (1971). 
\bibitem{Kosterlitz} J.M. Kosterlitz and D.J. Thouless, J. Phys. C {\bf 6}, 1181 (1973). 
\bibitem{Wang} Z. Wang, A. Kawakami, Y. Uzawa, and B. Komiyama, J. of Appl. Phys. {\bf 79}, 7837 (1996). 
\bibitem{Miki_IEEE} S. Miki, M. Fujiwara, M. Sasaki, and Z. Wang, 
IEEE Trans. Appl. Supercond. {\bf 17}, 285 (2007). 
\bibitem{Maksimova} G.M. Maksimova, Phys. Solid State {\bf 40}, 1607 (1998).
\bibitem{Kadin} A.M. Kadin, K. Epstein, and A.M. Goldman, Phys. Rev. B {\bf 27}, 6691 (1983). 
\bibitem{Mooij} J.E. Mooij, in {\it Percolation, Localization, and Superconductivity}, 
edited by A.M. Goldman and S.A. Wolf (Plenum, New York, 1984), p.325. 
\bibitem{Gorlova} I.G. Gorlova, S.G. Zybtsev, A.M. Nikitina, V.Y. Pokrovskii, V.N. Timofeev, 
and S. Aukkaravittayapun, JETP Lett. {\bf 68}, 216 (1998). 
\bibitem{Herbert} S.T. Herbert, Y. Jun, R.S. Newrock, C.J. Lobb, K. Ravindran, H.-K. Shin, D.B. Mast, 
and S. Elhamri, Phys. Rev. B {\bf 57}, 1154 (1998). 
\bibitem{Beasley} M.R. Beasley, J.E. Mooij, and T.P. Orlando, 
Phys. Rev. Lett. {\bf 42}, 1165 (1979). 
\bibitem{Jaeger} H.M. Jaeger, D.B. Haviland, B.G. Orr, and A.M. Goldman, 
Phys. Rev. B {\bf 40}, 182 (1989). 
\bibitem{lambda} We used an expression 
$\lambda(T) = \lambda(0){\left[(\Delta(T)/\Delta(0))
\tanh\left(\Delta(T)/2k_{B}T\right)\right]}^{-1/2}$ in Ref. 22 with 
the superconducting gap $\Delta(T) = \Delta(0)\tanh\left(1.74\sqrt{T_{c}/T - 1}\right)$. 
\bibitem{Tinkham} M. Tinkham, 
{\it Introduction to Superconductivity}, 2nd ed., (McGraw-Hill, New York, 1996). 
\end{thebibliography}
\end{document}